\documentclass[]{spie}  

\usepackage{amsmath,amsfonts,amssymb}
\usepackage{graphicx}
\usepackage{xcolor}
\usepackage{booktabs}
\usepackage{adjustbox}
\usepackage{tabularx,ragged2e}
\usepackage[colorlinks=true, allcolors=blue]{hyperref}
\DeclareUnicodeCharacter{03B3}{$\gamma$}

\title{A summary on an investigation of GAGG:Ce afterglow emission in the context of future space applications within the HERMES nanosatellite mission.}

\author[a,b,c]{Giuseppe~Dilillo}
\author[d,e]{Riccardo~Campana}
\author[c]{Nicola~Zampa}
\author[d,e]{Fabio~Fuschino}
\author[a,c]{Giovanni~Pauletta}
\author[h]{Irina~Rashevskaya}
\author[f]{Filippo~Ambrosino}
\author[a,c]{Marco~Baruzzo}
\author[a,c]{Diego~Cauz}
\author[a,c]{Daniela~Cirrincione}
\author[a,c]{Marco~Citossi}
\author[a,c]{Giovanni~Della~Casa}
\author[h]{Benedetto~Di~Ruzza}
\author[m,n]{Gábor~Galgóczi}
\author[d,e]{Claudio~Labanti}
\author[f,g]{Yuri~Evangelista}
\author[m,l,o]{Jakub~Ripa}
\author[a,c]{Andrea~Vacchi}
\author[i,h]{Francesco~Tommasino}
\author[h]{Enrico~Verroi}
\author[b]{Fabrizio~Fiore}

\affil[a]{University of Udine, Via delle Scienze 206, I-33100 Udine, Italy}
\affil[b]{INAF-OATs Via G.B. Tiepolo, 11, I-34143 Trieste}
\affil[c]{INFN sez. Trieste, Padriciano 99, I-34127 Trieste, Italy}
\affil[d]{INAF-OAS Bologna, Via Gobetti 101, I-40129 Bologna, Italy}
\affil[e]{INFN sez. Bologna, Viale Berti-Pichat 6/2, I-40127 Bologna, Italy}
\affil[f]{INAF-IAPS,Via del Fosso del Cavaliere 100, I-00133 Rome, Italy}
\affil[g]{INFN sez. Roma 2, Via della Ricerca Scientifica 1, I-00133 Rome, Italy}
\affil[h]{TIFPA-INFN, Via Sommarive 14, I-38123 Trento, Italy}
\affil[i]{Department of Physics, University of Trento, via Sommarive, 14 38123 Trento}
\affil[l]{Department of Theoretical Physics and Astrophysics, Faculty of Science, Masaryk University, Brno, Czech Republic}
\affil[m]{Eötvös Loránd University, Egyetem tér 1-3, 1053 Budapest, Hungary}
\affil[n]{Hungarian Academy of Sciences, Wigner Research Centre for Physics,1525 Budapest 114, Hungary}
\affil[o]{Astronomical Institute of Charles University,  V Holešovičkách 747/2, CZ-18000 Prague 8, Czech Republic}

\begin{document} 
\maketitle

\begin{abstract}
GAGG:Ce (Cerium-doped Gadolinium Aluminium Gallium Garnet) is a promising new scintillator crystal. A wide array of interesting features --- such as high light output, fast decay times, almost non-existent intrinsic background and robustness --- make GAGG:Ce an interesting candidate as a component of new space-based gamma-ray detectors.
As a consequence of its novelty, literature on GAGG:Ce is still lacking on points crucial to its applicability in space missions. In particular, GAGG:Ce is characterized by unusually high and long-lasting delayed luminescence. This \textit{afterglow} emission can be stimulated by the interactions between the scintillator and the particles of the near-Earth radiation environment. By contributing to the noise, it will impact the detector performance to some degree.
In this manuscript we summarize the results of an irradiation campaign of GAGG:Ce crystals with protons, conducted in the framework of the HERMES-TP/SP (High Energy Rapid Modular Ensemble of Satellites - Technological and Scientific Pathfinder) mission. A GAGG:Ce sample was irradiated with 70 MeV protons, at doses equivalent to those expected in equatorial and sun-synchronous Low-Earth orbits over orbital periods spanning 6 months to 10 years, time lapses representative of satellite lifetimes.
We introduce a new model of GAGG:Ce afterglow emission able to fully capture our observations. Results are applied to the HERMES-TP/SP scenario, aiming at an upper-bound estimate of the detector performance degradation due to the afterglow emission expected from the interaction between the scintillator and the near-Earth radiation environment.
\end{abstract}

\section{Introduction}

HERMES-Technologic and Scientific Pathfinder (HERMES-TP/SP) is a constellation of six 3U nanosatellites hosting simple but innovative X-ray detectors for the monitoring of Cosmic High Energy transients such as Gamma Ray Bursts and the electromagnetic counterparts of Gravitational Wave Events \cite{fiore20}. The main objective of HERMES-TP/SP is to prove that accurate position of high energy cosmic transients can be obtained using miniaturized hardware, with cost at least one order of magnitude smaller than that of conventional scientific space observatories and development time as short as a few years.
The main goals of the projects are: 1) join the multimessenger revolution by providing a first mini-constellation for GRB localizations with a total of six units, performing a first experiment of GRB triangulation with miniaturized instrumentation; 2) develop miniaturized payload technology for breakthrough science; 3) demonstrate COTS applicability to challenging missions, contribute to Space 4.0 goals; push and prepare for high reliability large constellations.

The HERMES-TP project is funded by the Italian Ministry for education, university and research and the Italian Space Agency. The HERMES-SP project is funded by the European Union’s Horizon 2020 Research and Innovation Programme under Grant Agreement No. 821896. 
The constellation should be tested in orbit in 2022. 
HERMES-TP/SP is intrinsically a modular experiment that can be naturally expanded to provide a global, sensitive all sky monitor for high energy transients.

\begin{figure}[h!]
\centering
\includegraphics[width=0.7\textwidth]{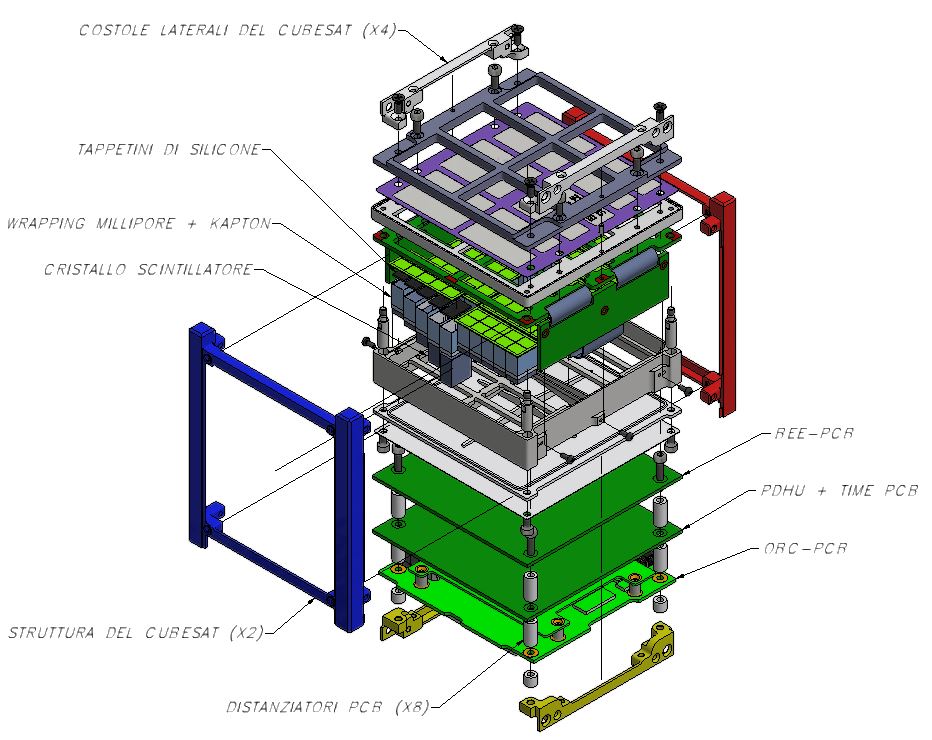}
\caption{Exploded view of the HERMES payload} 
\label{fig:esploso}
\end{figure}

The foreseen detector for this experiment employ the solid-state Silicon Drift Detectors (SDD) developed by INFN and FBK in the framework of the ReDSoX Collaboration\footnote{\url{http://redsox.iasfbo.inaf.it}}. These devices, being sensitive to both X-ray and optical photons, and characterised by a very low intrinsic electronic noise, can be exploited both as direct X-ray detectors and as photo detectors for the scintillation light produced by the absorption of a gamma-ray in an inorganic scintillator crystal \cite{Campana:2017jls}. This allows for the realisation of a single, compact experiment with a sensitivity band from a few keV to a few MeVs for X and gamma-rays, and with a high temporal resolution ($<1\,\mu$s)\cite{fuschino2019}. Figure~\ref{fig:esploso} shows a sketch of the proposed payload (which will fit in a 3U CubeSat platform). 

Since the nanosatellites will be launched as secondary payloads in a low-Earth orbit, they are subject to a relatively large amount of high-energy radiation fluxes (mostly cosmic-rays and geomagnetically trapped protons). This results in a degradation of the performance of the scintillator crystals (due to e.g. afterglow, activation and creation of additional luminescence centers) and of the silicon detectors (mainly due to the Non-Ionizing Energy Loss radiation damage, leading to an increase of the leakage current).

The aim of the beam test campaign at the Protontherapy centre was to gain insight on the behaviour of both the proposed scintillator crystal, when exposed to proton doses representative of typical values encountered on orbit during the whole operative life (at least 1–2 years).

\section{Background and aims}

For the HERMES detectors, the choice of the optimal scintillator material required a careful evaluation of several factors:
\begin{itemize}
\item Maximisation of the light output (photons per unit of absorbed energy).
\item Non-hygroscopicity of the crystal.
\item High density and average atomic number (stopping power).
\item Good radiation-resistance properties.
\item Low light emission characteristic time.
\end{itemize}
Therefore, the choice has fallen on a relatively recent material, the Cerium-doped Gadolinium-Aluminium-Gallium Garnet (Gd$_3$Al$_2$Ga$_3$O$_{12}$ or GAGG:Ce), developed firstly in Japan around 2010, and commercially available since 2014. This material has a high intrinsic light output ($\sim$50,000~ph/MeV), no intrinsic background, no  hygroscopicity, a fast radiation decay time of $\sim$90~ns, a high density ($6.63$~g/cm$^3$), a peak light emission at $520$~nm and an effective mean atomic number of $54.4$. All these characteristics make this material very suitable for the HERMES application.

Since GAGG is a relatively new material, it has not yet been extensively investigated with respect to radiation resistance and performance after irradiation, although the published results are very encouraging\cite{Yoneyama_2018}. These literature results (performed with fixed proton doses up to $10–100$~krad and $^{60}$Co gamma-ray doses of $100$~krad) showed that, compared to other scintillator materials largely used in the past years in space-borne experiments for gamma-ray astronomy (e.g. BGO or CsI), GAGG has a very good performance, i.e. a very low activation background (down to 2 orders of magnitude lower than BGO), and an inferior light output degradation with accumulated dose. However, an issue is the non-negligible amount of afterglow observed after a $100$~krad $^{60}$Co irradiation: the crystals exhibits a phenomenon of phosphorescence on long time scale, that will lead to a slight transitory worsening of the energy resolution and light output.

The first objective of this campaign was therefore to further investigate these results by a careful evaluation of the afterglow phenomenon and the variations in optical properties of the scintillator with proton doses and at increasing dose steps, representative of the actual in-orbit radiation environment foreseen for HERMES. In the sections that follow we will  outline the experimental procedure and summarize the results from the data analysis. A more complete paper on the subject is currently in preparation\cite{dilillo20}.

\section{Experimental procedure and setup}
\label{sec:expout}
The characterization was carried on at the TIFPA-APSS experimental area of the Trento Proton Therapy Center, with irradiated doses chosen to be representative of typical fluxes encountered in the foreseen HERMES low-Earth orbit, at a beam energy of 70 MeV and using several irradiation steps (cf. Table~\ref{tab:FULLdoses}). 
A teflon-wrapped GAGG:Ce scintillator crystal---dimensions $3 \times 1 \times 1$ cm$^3$---was housed in a lightproof metal case and optically coupled to a pair of photomultiplier tubes (PMT) through an optical waveguide a few cm long, to ensure the PMT shielding from the proton beam. The afterglow effects were measured immediately after (from $60$~s up to $\sim 1000$~s) the irradiation, by monitoring the amount of current flowing through the PMT anode (proportional to the scintillation light emitted by the crystal and collected by the PMT) by means of a picoammeter and a data logger.
Scintillator optical properties such as characteristic emission time, light output and energy resolution were then evaluated by acquiring spectra of single proton events by means of a waveform digitizer, and comparing the results with those obtained before the irradiation.
The anode signal of one of the two PMTs was measured by a Keithley 6487 picoammeter. The signals from the last dynode of both PMTs were brought to a discriminator unit where they were split and redirected to a counter and a multi-channel digitizer for waveform acquisition.

\definecolor{color0}{HTML}{000000}

\definecolor{color1}{HTML}{5c4ccf} 
\definecolor{color2}{HTML}{a3d4ff} 
\definecolor{color3}{HTML}{14b385} 
\definecolor{color4}{HTML}{a56bbf} 
\definecolor{color5}{HTML}{fa9507} 
\definecolor{color6}{HTML}{e84031}

\begin{table}
\centering
\begin{adjustbox}{angle=0,max width=\textwidth}  
\begin{tabularx}{0.9\textheight}{@{}>{\RaggedRight}X*{8}{ c@{\hskip.1cm}} @{}}

\centering

 & \multicolumn{6}{c}{} \\
 & {} & {} & {} & {} & {} & {} & {Measurements}& {} \\ 
\cmidrule(l){7-9}
& {} & { \ Crystal} & { \ Irr. Duration [s]} & { \ Dose$^{[1]}$ [p]} & { \ E.O.P.$^{[2]}$} & { \ Current} & { \ Counts}& { \ Waveform} \\

\midrule
 2019.01.30 19:59 
& \textcolor{color1}{$\bullet$} &  J2 &  $90$ s &  $(1.19 \pm 0.23)\times10^{8}$ &  $1$y EQ & $\checkmark$ & $\checkmark$ & $\checkmark$ \\
 2019.01.30 20:22
& \textcolor{color2}{$\bullet$} &  J2 &  $90$ s & $(1.20 \pm 0.23)\times10^{8}$ &  $2$y EQ & $\checkmark$ & $\checkmark$ & $\checkmark$ \\
 2019.01.30 20:48
& \textcolor{color3}{$\bullet$} &  J2 &  $270$ s & $(3.56 \pm 0.69)\times10^{8}$ &  $5$y EQ & $\checkmark$ & $\checkmark$ & $\checkmark$ \\
2019.01.30 21:12 
& \textcolor{color4}{$\bullet$} &  J2 &  $100$ s & $(1.37 \pm 0.26)\times10^{9}$ &  $10$y EQ & $\checkmark$ & $\checkmark$ & $\checkmark$ \\
 2019.01.30 21:37
& \textcolor{color5}{$\bullet$} &  J2 &  $144$ s & $(1.95 \pm 0.37)\times10^{9}$ &  $2$y SSO & $\checkmark$ & $\checkmark$& $\checkmark$ \\
 2019.01.30 22:02
& \textcolor{color6}{$\bullet$} &  J2 &  $115$ s & $(1.52 \pm 0.29)\times10^{10}$ &  $10$y SSO & $\checkmark$ & $\checkmark$& $\checkmark$ \\
\midrule

\end{tabularx}
\end{adjustbox}

\caption{
Detailed table of the irradiation runs. Each row represents an irradiation step.\newline Runs are identified by current log start time and color coded as in article body.  Temperatures in range $21 \pm 0.5 \,^\circ$C. \newline [1] : Estimated from GEANT4 simulation of a proton beam irradiation of GAGG:Ce crystal. The $70$ MeV proton beam is modelled with a Gaussian shape, having a width of $\sigma = 6.9$ mm, according to the beam characteristics at the isocenter point \cite{TPTC}.  The crystal is placed at the beam center, and possible positioning errors have been taken into account. The total number of protons simulated has been generated according to the flux measured by the beam monitor thus calculating the total energy deposit and resulting dose.\newline
[2] : Equivalent Orbital Period. The reported values equal the flux of proton with energies $> 0.1$ MeV expected in equatorial (EQ) or Sun-Synchronous (SSO) orbits according to AP8MIN models (cf. Section~\ref{sec:leakage}) integrated over time periods comparable to the expected spacecraft orbital lifespan. 
}

\label{tab:FULLdoses}

\end{table}

\section{Results}

\subsection{Afterglow model and irradiation data analysis}
\label{sec:afterglow}

Long-lived afterglow emission in scintillators is due to presence of impurity sites or defects within the crystal lattice.
Arriving at the impurity sites, charge carriers (electrons, holes) can create excited configurations whose transition to ground state is forbidden\cite{knoll}. At later times, charge carriers can escape these sites by different processes (e.g. to the conduction band by thermal energy absorption or to nearby recombination centers by direct or thermally assisted tunneling \cite{2006:huntley}). Ultimately all of the charge carriers recombine, giving rise to luminescence. 

Different scintillators displays different afterglow characteristics, varying from almost non-existent to a very long and intense emission \cite{Yoneyama_2018}. GAGG:Ce has been known to belong to the end of the spectrum showing intense emission lasting up to several days \cite{Yoneyama_2018}. Afterglow mitigation techniques, such as Mg-codoping, proved successful at cost of diminished light-yield\cite{LUCCHINI2016176}.

We propose a semi-empirical model in which \textit{i.} afterglow emission results from the release of charge carriers trapped in metastable states with few different, discrete lifetimes; \textit{ii.} the density of a particular trap species can change linearly with the irradiation dose. Making use of such a model one can determine the number of charge carriers in a particular trap species during each  irradiation step. Once the irradiation ceases, the number of trapped charge carriers decreases exponentially with time.

The PMT anode current we observe will be proportional to the time derivative of the total number of trapped charges, $I(t) \propto \frac{dN}{dt}$, the proportionality constant depending from a number of quantities measurable through calibration, namely the PMT gain and quantum efficiency, and the photon transport efficiency from the crystal to the photocathode. 
Through a fitting procedure, we derive the model parameters i.e., $\tau_i$, trap emission time constants, $n_i$, the density of charge carriers trapped for particle of incident radiation and  $\Delta  n_i$, the the total variation in trap species density at the end of the last irradiation.  
In this procedure we start by considering only the data from the first few irradiation and measurement runs, then progressively add subsequent measurements. At each step we use the partial model to check the accuracy of the predictions for the following irradiations. The model is then progressively adjusted to successfully fit the whole dataset.

We find that the smallest number of trap species needed to resolve patterns in fit residuals of the last measurement run is 7. Models with more trap species results in worse estimates of best fit parameters while providing no enhancements in the residuals.

\begin{figure}[htb]
\centering
\includegraphics[width=0.8\textwidth]{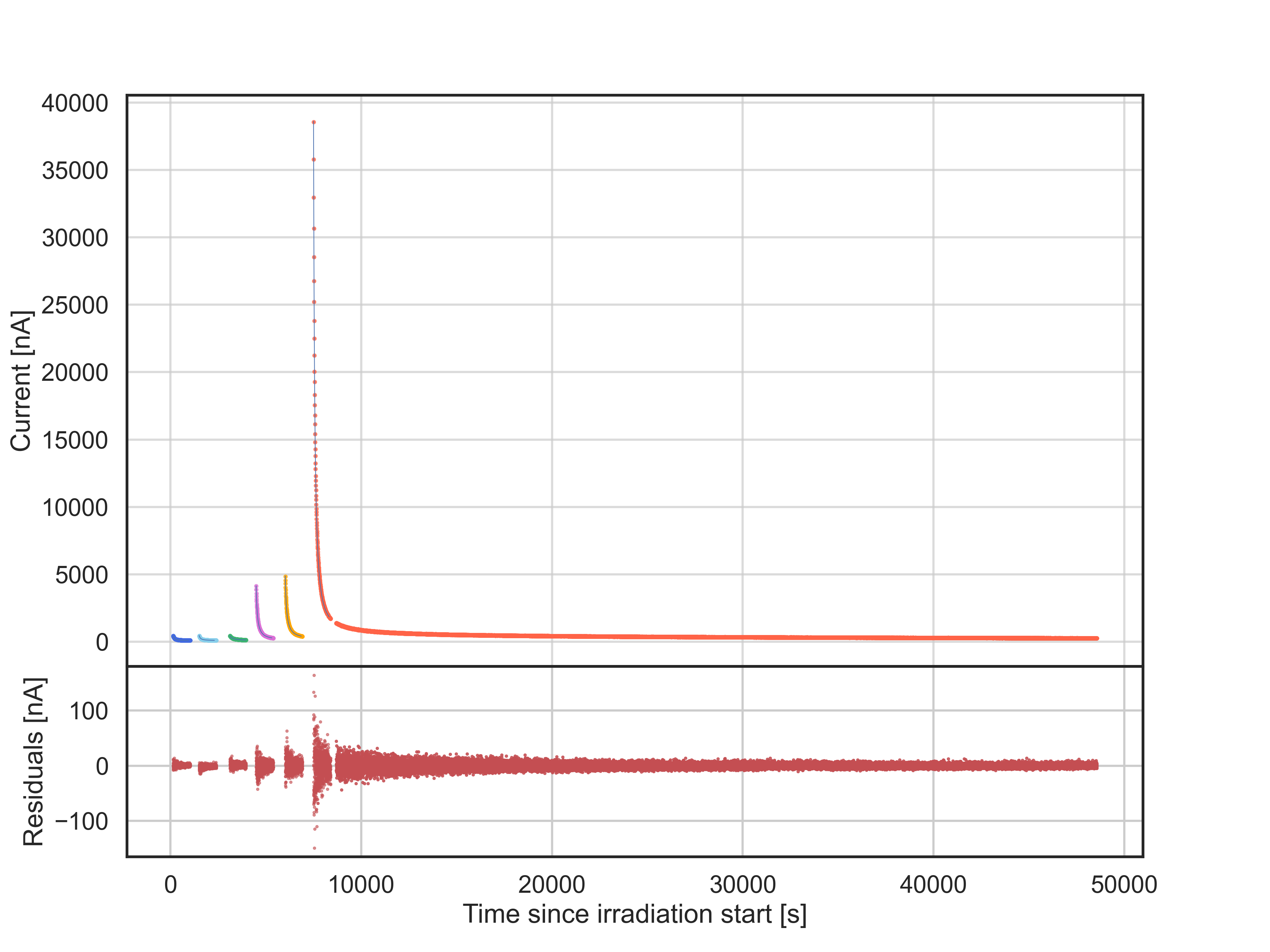}
\caption{The dataset fitted to the model outlined in Section~\ref{sec:afterglow}. Measurements, best fit curves and residuals are shown. Parameters estimates as in Table~\ref{tab:finfit_nt}. Temperatures are in range $21 \pm 0.5 \, ^\circ$C} 
\label{fig:finfit}
\end{figure}

Fit results are plotted in Figure~\ref{fig:finfit}. Corresponding best parameters estimates are reported in Table~\ref{tab:finfit_nt}. 
For two of the trap species we find the probability of capture to grow with the accumulated dose, starting from initially negligible values.
At variance, one of the trap species ceases to capture carriers during the last irradiation. 
For the latter, an emission component is still found in the last measurement. We believe to be reasonable to suppose that such an effect could stem from saturation of the trap species: at some point during irradiation, the traps of this species are all occupied and can no longer capture new charges.
Beside trap saturation, other processes that could explain such phenomena are variations in the traps average lifetime, formations of new traps or reduction in the number of existing traps, dose dependent formation of non-radiative centers and other dose-dependent non-linear effects.
From the present data we cannot take the investigation of these phenomena further or even exclude the possibility that such effects could arise as a consequence of the experimental procedure.

%
%
%
%
%
%

\begin{table}[]
\centering
    \begin{minipage}{.6\textwidth}
      \centering
       \resizebox{\textwidth}{!}{%
		\begin{tabular}{llllll}
		\toprule
		$\tau_i$ [$s$] & $\sigma_\tau$ [$s$] & $n_i$ [$cm^{-3}$]& $\sigma_n$ [$cm^{-3}$] & $\Delta n$ [$cm^{-3}$]& $\sigma_{\Delta n}$ [$cm^{-3}$]\\
		\midrule
		
		$25.32$ & $0.41$ & $196.0$ & $3.5$ & $69.0$ &  $4.2$\\
		$68.97$ & $1.00$ & $129.4$ & $1.7$ & $-27.9$ & $1.2$\\
		$188.8$ & $1.6$ & $94.8$ & $1.2$ & $-19.3$ & $1.2$\\
		$656.7$ & $7.5$ & $63.9$ & $2.0$ & $-46.2$ & $3.2$\\
		$2103$ & $32$ & $0.0$ & $5.4$ & $47.2$ & $9.6$\\
		$8021$ & $98$ & $152.2$ & $2.6$ & $-222.2$ & $6.9$\\
		$6.880\cdot 10^4$ & $437.6$ & $0.0$ & $48.0$ & $1669$ & $124$\\

		\bottomrule
		\end{tabular}}
		\caption{Traps properties as estimated from fit to the model of Section~\ref{sec:afterglow}. Afterglow data from a GAGG:Ce sample at temperature $21 \pm 0.5 \, ^\circ$C.}
		\label{tab:finfit_nt}
    \end{minipage}
\end{table}

\subsection{Impact of GAGG:Ce afterglow on silicon drift detectors}
\label{sec:leakage}
The first HERMES spacecrafts are expected to fly on a LEO, near-equatorial orbit. These units are expected to be operative for a minimum of $2$ years.
We expect afterglow emission to be stimulated in orbit by interaction between the scintillator and particles trapped in the Van Allen radiation belt. Having no intrinsic gain, SDD cells coupled to a scintillator will not be able to resolve the dim afterglow photons into full-blown signals.
Still, the afterglow will contribute to noise in a way similar to the detector leakage current causing a degradation of the detector performance. \\ Through the model outlined in Section~\ref{sec:afterglow} we are now able to predict the impact of GAGG:Ce afterglow emission in terms of equivalent leakage current. To estimate the trapped particle fluxes along the orbit we use the IRENE (International Radiation Environment Near Earth) AE9/AP9 models \cite{irenemodel}. These models allows to compute proton and electron orbital fluxes in space. The previous AE8/AP8 versions were developed by NASA, are regarded as the industry standard for radiation belt modeling, and are available in the MIN and MAX variants accounting for minimum or maximum of solar activity \cite{misc:esa_models}. The AE9/AP9 models ---which are built upon much more recent trapped radiation observations--- are expected to  replace their predecessors in the near future. 

Assuming the scintillator to be irradiated at constant average rate for periods of times equal to the simulation timesteps ($10$ s) and considering the flux values calculated with AE9/AP9 along $30$ days of orbit, we can estimate the number of trapped charges $N(t)$ at some orbital time using the model parameterization of Section~\ref{sec:afterglow}.

We are interested in an upper-bound estimate of the noise to be expected from afterglow. Hence we analyze the following worst-case scenario: for each trap species we consider the maximum capture capacity between the values expected at mission start and mission end. The expected value of leakage current is $I_L(t) = - e f \epsilon \frac{dN}{dt}$, where $N$ is the number of trapped electrons at a given time, $\epsilon$ indicates the quantum efficiency of the SDD, $f$ is the crystal to SDD photon transport efficiency and $e$ is the elementary charge. 

\begin{figure}
\centering
\includegraphics[width=0.8\textwidth]{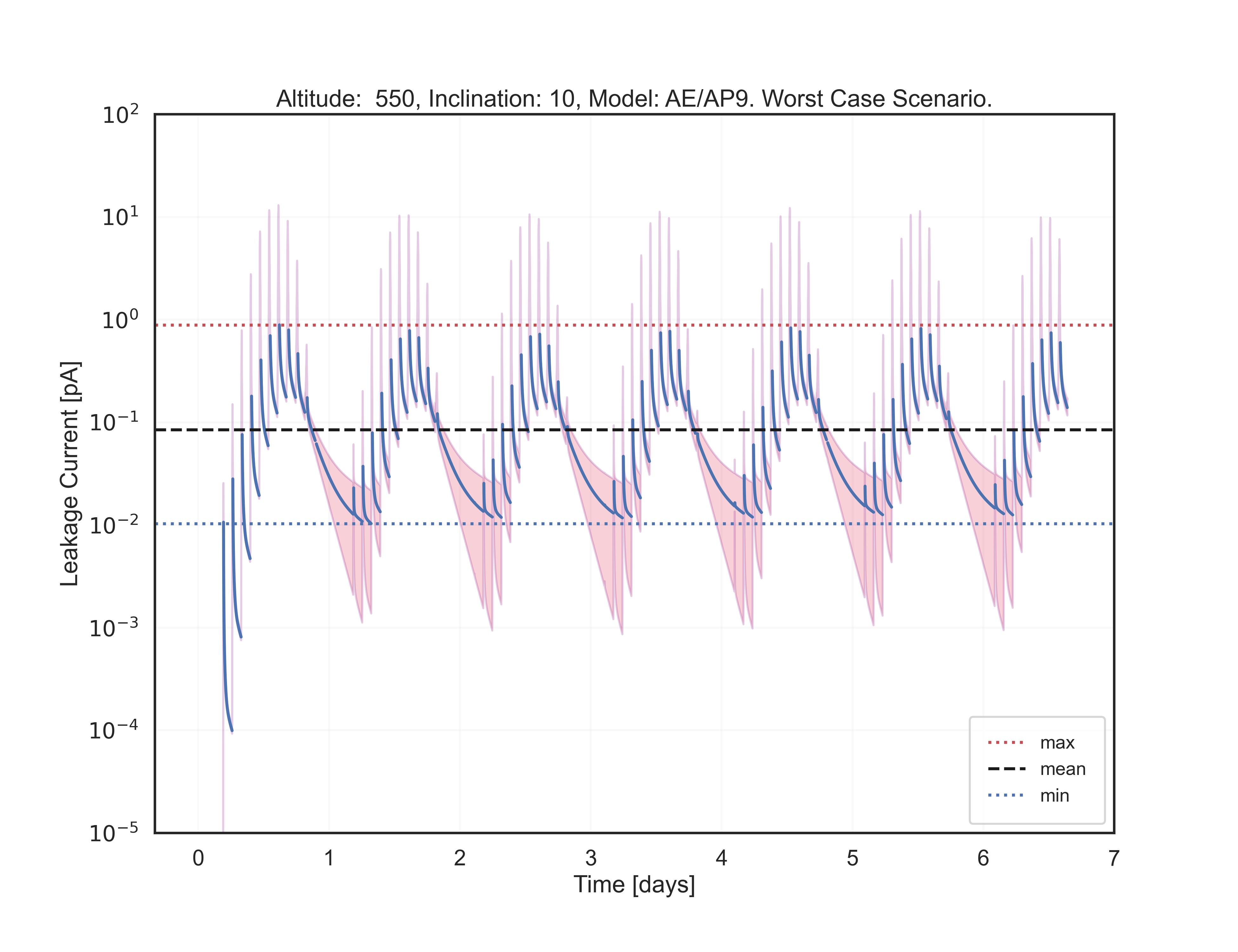}
\caption{Estimated worst-case leakage current of a SDD cell with dimensions $6.94 \times 6.05$ mm$^2$ resulting from GAGG:Ce afterglow emission induced by irradiation of a $12.1 \times 6.94 \times 14.50$ mm$^3$ scintillator at temperature of $21 \pm 0.5 \, ^\circ$C. The scintillator is completely  shielded from radiation on one of the smaller face. Orbital populations of protons and electrons are modelled through  AE9/AP9 packages for an orbit with 550~km altitude and  $10^\circ$ inclination orbit over $\sim$7 days period (100 orbits). 
Values during transits over SAA are not reported.
The uncertainty region is represented as a shaded band. Reported values of minimum, mean and maximum leakage current are calculated starting from $24$ hours of orbital lifetime. Afterglow model as outlined in Section~\ref{sec:afterglow}. The model parameters are reported in Table~\ref{tab:finfit_nt}.} 
\label{fig:leakage_worstcase}
\end{figure}

In Figure~\ref{fig:leakage_worstcase} we report our worst-case estimate of the  leakage current resulting from afterglow emission of a GAGG:Ce sample with HERMES detector dimensions, as expected over $100$ orbits at altitude $550$ km and inclination $10^\circ$. 
The reported minimum, mean and maximum values of leakage current are calculated starting at $24$ hours of orbital lifetime. 
We remark that the data on which the model from Section~\ref{sec:afterglow} was built were gathered starting from a minute after the end of the irradiations. It follows that the model is expected to fail inside and up to one minute following a passage over trapped radiation regions. For our purposes this is not a problem since in these regions the HERMES instruments will be turned off. For this reason, the current values expected during such transits are omitted in Figure~\ref{fig:leakage_worstcase}.

HERMES low-noise front-end electronics (FEE) will be able to grant nominal performance up to $\sim$100~pA of leakage current, a value which is well above two order of magnitude from the estimated maximum \cite{fuschino_grassi}. Despite this fact, increases in leakage current will still result in a worsening of the detector energy resolution.

Since the first six HERMES Technological Pathfinder units will be launched in near-equatorial orbits, we expect the afterglow impact on payload performance to be small.
However, when the HERMES fleet will be enlarged to host spacecrafts in orbits at higher inclinations, the impact of afterglow on detector performance will need further, more accurate investigations.

In fact, the flux of trapped radiation varies greatly with the inclination of the orbit. As a consequence, we expect the intensity of the afterglow emission to change accordingly. 
Repeating our worst-case evaluation for an orbit with altitude $550$ km and $50$ degrees inclination we find the leakage current from afterglow emission to exceed the FEE's nominal performance limit of $100$ pA. However one must note that producing these results our assumptions about the afterglow and trapped particles models \cite{ripa20}, shielding, kinetic-ionization conversion inside the scintillator and SDD quantum efficiency have all been very conservative. Shielding in particular is expected to play a substantial effect in moderating the afterglow component due to interactions between scintillator and trapped belt radiation. An accurate estimate of the degradation in detector performances due to GAGG:Ce afterglow emission for spacecrafts in orbits at higher inclination cannot be separated from a careful description of the payload and spacecraft itself, something which goes beyond the scope of this work.

\section*{ACKNOWLEDGMENTS}
This project has received funding from the European Union Horizon 2020 Research and Innovation Framework Programme under grant agreement HERMES-Scientific Pathfinder n. 821896 and from ASI-INAF Accordo Attuativo HERMES Technologic Pathfinder n. n. 2018-10-HH.0. We also thank the TIFPA staff for their precious assistance and the ReDSoX collaboration. 

\bibliography{report} 
\bibliographystyle{spiebib} 

\end{document}